# Wavelength division multiplexed and double-port pumped time-bin entangled photon pair generation using Si ring resonator

**MIKIO FUJIWARA,**[1,*] **RYOTA WAKABAYASHI,**[1] **MASAHIDE SASAKI,**[1] **AND MASAHIRO TAKEOKA**[1]

[1]*Quantum ICT Advanced Development Center National Institute of Information and Communications Technology (NICT), 4-2-1 Nukui-kita, Koganei, Tokyo 184-8795, Japan*
*\*fujiwara@nict.go.jp*

**Abstract:** We report a wavelength division multiplexed time-bin entangled photon pair source in telecom wavelength using a 10 μm radius Si ring resonator. This compact resonator has two add ports and two drop ports. By pumping one add port by a continuous laser, we demonstrate an efficient generation of two-wavelength division multiplexed time-bin entangled photon pairs in the telecom C-band, which come out of one drop port, and are then split into the signal and idler photons via a wavelength filter. The resonator structure enhances four wave mixing for pair generation. Moreover, we demonstrate the double-port pumping where two counter propagating pump light are injected to generate entanglement from the two drop ports simultaneously. We successfully observe the highly entangled outputs from both two drop ports. Surprisingly, the count rate at each drop port is even increased by twice as that of the single-port pumping. Possible mechanisms of this observation are discussed. Our technique allows for the efficient use of the Si ring resonator, and widens its functionality for variety of applications.

**OCIS codes:** (270.5568) Quantum cryptography; (060.5565) Quantum communication.

## 1. Introduction

Establishing secure data transmission is an urgent issue for consumers, businesses, and governments because highly confidential data such as genomic data or diplomatic documents are transferred through internet. One approach toward the goal is to develop a so-called "post quantum cryptography". This is based on computational complexity as conventional cryptography, but it is expected to be secure against even deciphering by quantum computer [1]. Various protocols have been considered and the estimation of their security level is ongoing [2]. Another approach is quantum key distribution (QKD) [3-5]. It can guarantee the information-theoretic security by laws of physics, which is already proved to be secure against any computational attacks including quantum computing.

To make QKD a commercially available technology, it is highly desirable to have small and low cost quantum devices. Integrated photonics technology can realize the device downsizing. Especially, Si photonics is a promising approach [6-9] because it has compatibility with complementary metal oxide semiconductor (CMOS) process. It is useful not only for the conventional QKD with weak coherent laser sources, but also for the next generation QKD based on entangled photon sources, since Si waveguides and resonators can have large third-order nonlinear susceptibility. The large nonlinearity provides high generation rate of photon pairs by spontaneous four wave mixing (SFWM) [10-14]. Entangled photon pair generation using Si ring resonators was theoretically studied in [15], and polarization entangled photon pairs were generated from a Si ring resonator [16].

Since QKD is usually linked through fibers, time-bin [17] entangled photon pairs are most preferred due to its robustness in fiber propagation. In our previous work [18], we succeeded in generating time-bin entangled photon pairs at a pair of signal and idler wavelengths in the telecom C-band using a 7 μm radius Si ring. Thanks to the ring resonance structure, the photon pairs are generated efficiently with narrowband widths. In addition, free spectral range (FSR) is flexibly selected by adjusting a radius of the resonator, which allows for wavelength tunability as well as channel multiplexing. Therefore Si ring resonator offers a promising light source which suits to compact entanglement based QKD implementation. Another result on time-bin entangled photon pair generation by Si ring has been reported [19].

In this paper, we extend our previous scheme [18] by two directions. One is wavelength division multiplexing in the telecom C-band at one drop port, and the other is duplex generation from the two drop ports by the double-port pumping. For the former purpose, we employ a Si ring with a larger radius, 10 μm, than the previous one with 7 μm radius and tune the pump laser and the resonant peak of the ring properly to generate two-wavelength division multiplexed entangled photon pairs in the telecom C-band. The signal and idler photons at each channel from one drop port are split into two fibers via a wavelength filter. . For the latter purpose, we develop the double-port pumping technique where the ring is pumped in two counter propagating directions simultaneously. We successfully observe the entangled outputs from both two drop ports. Surprisingly, the count rate at each drop port is larger than that of the single-port pumping with the same pump power, namely increases almost by twice. Possible mechanisms of the count rate enhancement are discussed.

Note that after completing this work, we were aware of the related work on the wavelength multiplexed generation of time-bin entanglement from a Si ring [20]. In [20], entangled photon pairs are generated from various wavelengths tuned to be the ITU grid at the telecom C-band. The ring has 60 μm radius which is larger than ours and the double-port pumping has not been examined.

The paper is organized as follows. In section 2, we show the design of the Si ring resonator, and describe the experimental setup for measuring time-bin entanglement of photon pairs. The results on wavelength multiplexed generation of time-bin entanglement and the double-port pumping experiment are shown in section 3. Finally, we summarize our results and conclude the paper in section 4.

## 2. Design of the photon pair generator and experimental setup

The Si ring resonator is fabricated by CMOS-compatible processes (made by NTT-AT Corporation). The conceptual view and photograph of the resonator is shown in Fig. 1(a).The Si ring resonator with 10 μm radius horizontally couples to two waveguides across 350 nm gap. The cross section of the ring and straight line waveguides is a square of 400 nm width and 220 nm thickness. Pump light is injected into the resonator via an add port in the input waveguide and the photon pair is output through a drop port in the other waveguide. Every port of the two waveguides is connected to a single mode fiber through a spot size convertor. The transmission spectrum of the ring resonator is shown in Fig.2. A broad band amplified spontaneous emission source and an optical spectrum analyzer [ADVANTEST Q8384] are used in this measurement. The resonator is temperature controlled by a Peltier device shown in Fig. 1(b) and set at 28.907±0.001°C. The quality factors of these resonant peaks are 6000-10000.

The wavelength of the pump laser is adjusted at the peak of the resonance which excites SFWM. The first neighbor pair and the second neighbor pair, almost belong to telecom C-band, are used for the generation of the wavelength division multiplexed time-bin entanglement (see [Fig. 2]). The setup for the pump source is illustrated in Fig. 4(a) as the "single-port photon pair generator" (surrounded by the dotted line). In order to eliminate the noise in the pump light, an optical narrow bandpass filter (2 nm) is put in front of the wavelength tunable laser (Agilent 81980A).

A fiber in-line module consisting of a polarizer, a half wave plate (HWP), and a quarter wave plate (QWP) is used to adjust the pump polarization to the quasi-TE mode in the ring.
Generated photon pairs in the resonator exit at the drop port of the waveguide. A Fiber Bragg Grating (FBG) filter is used to cut the pump light. The signal and idler photons are divided by a dense wavelength division multiplexing (DWDM) filter, and are sent to two fiber channels separately. For both signal and idler channels, optical narrow bandpass filters (2 nm) are set to select the first or second photon pairs. Before the time-bin entanglement experiment, the outputs from the bandpass filters are directly connected to the superconducting single photon detectors (SSPDs) to check the coincidence counts [21, 22]. The polarization of the photons incident into each SSPD is controlled by a fiber polarization controller to compensate the polarization dependence of the detection efficiencies of the SSPD. Electrical pulses from the SSPDs are input to the time interval analyzer (TIA [Hydra harp 400]) for analyzing coincidence counts and the time-bin entanglement.

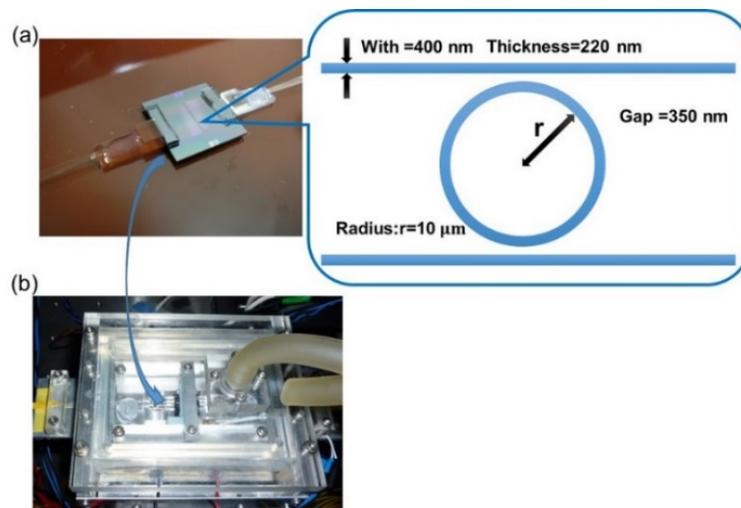

Fig. 1. (a) Photograph of the Si ring resonator with a fiber array. Inset, the dimension of the resonator is described. (b) Photograph of the cooling box. The ring resonator is installed in the center of the box.

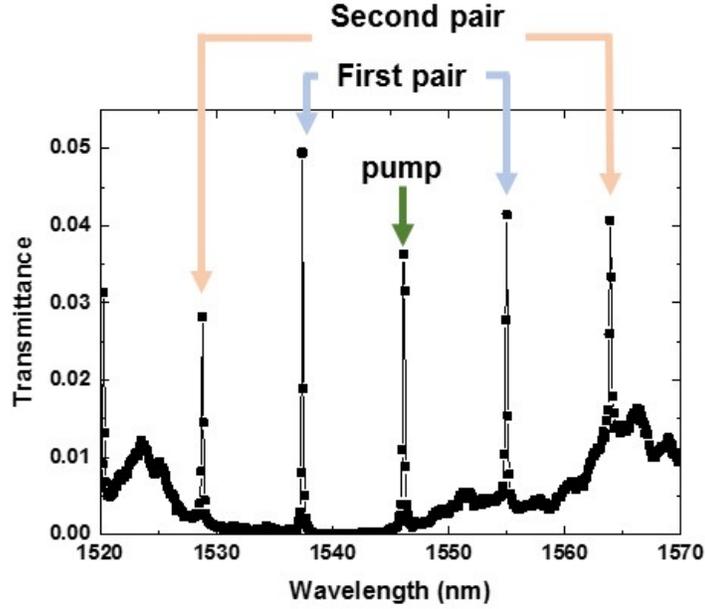

Fig. 2. Resonant peaks of the Si ring resonator of 10 μm radius at 28.907 degree centigrade. The wavelength of a pump laser is set at 1546.0593 nm. Photon pairs of the first and second are used in time-bin entanglement experiments.

Coincidence count of the second pair is shown in Fig. 3. The power of the pump light at 1546.0593±0.001 nm is set to be 0.5 mW. The time window for coincidence counts is 64 ps. We obtain the coincidence to accidental ratio (CAR) as more than 350. The transmission losses including detection losses are estimated to be 28dB for the signal and 29dB for the idler, respectively.

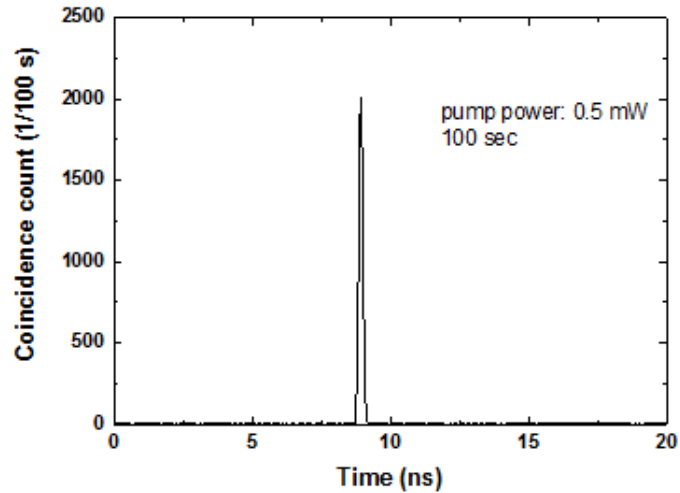

Fig. 3. Coincidence histogram of the second pair. The pump power is set -3.0 dBm (0.5 mW). The operation temperature of the Si ring resonator is 28.907 °C. The time window for one data point is set 64 ps.

## 3. Wavelength division multiplexed time-bin entanglement and double-port pumped time-bin entangled photon pair generation

A full setup of the time-bin entanglement generation is shown in Fig. 4(a). The time-bin entangled state [20] with a delay time of $\tau$ is written as [18];

$$\frac{1}{\sqrt{2}}\left\{\hat{a}^\dagger(t)\hat{b}^\dagger(t)+e^{i\theta}\hat{a}^\dagger(t+\tau)\hat{b}^\dagger(t+\tau)\right\}|0_a,0_b\rangle. \qquad (1)$$

Here $\hat{a}^\dagger(t)$ and $\hat{b}^\dagger(t)$ denote the creation operators of the signal and idler photons, and $|0_a,0_b\rangle$ denotes the vacuum state of the signal and idler mode. $\theta$ corresponds to the relative phase difference between two pulses. In our case, the continuous wave (CW) laser is used to pump the Si ring resonator, and photon pairs mathematically written as $\hat{a}^\dagger(t)\hat{b}^\dagger(t)|0_a,0_b\rangle$ are generated in a CW beam in the asymmetric Mach-Zehnder Interferometer (AMZI) with the delay time of $\tau$ at both signal and idler channels.. These AMZIs are fabricated in the planar lightwave circuit (PLC) [23] based on silica waveguide technology. The PLC has two-inputs and four-outputs and is originally designed as a photon decoder for QKD. These 2×4 AMZIs allow one to define the time-bin entanglement with the delay $\tau$ as well as randomly selecting their measurement basis from the X-basis or the Z-basis. Measurement of the Z-basis corresponds to the correlation measurement of the arrival time of the photon pairs, and of the X-basis is used to measure two photon interference. Basis selection can be made by the photons themselves in the AMZIs, which is intrinsically random. The PLC-AMZIs can operate in a polarization insensitive way by carefully adjusting the birefringence with precise temperature control. The entangled state measured at the output ports X (X0, X1, X'0 and X'1) is written as;

$$|\psi\rangle_{\text{final(X-basis)}} = \frac{1}{2\sqrt{2}}\begin{bmatrix}\left(1+e^{i(\theta_1+\theta_2+\theta(t-\tau)-\theta(t))}\right)\hat{a}^\dagger(t)_{X0}\hat{b}^\dagger(t)_{X'0} \\ +\left(1+e^{i(\theta_1+\theta_2+\theta(t-\tau)-\theta(t))}\right)\hat{a}^\dagger(t)_{X1}\hat{b}^\dagger(t)_{X'1} \\ +\left(1-e^{i(\theta_1+\theta_2+\theta(t-\tau)-\theta(t))}\right)\hat{a}^\dagger(t)_{X0}\hat{b}^\dagger(t)_{X'1} \\ +\left(1-e^{i(\theta_1+\theta_2+\theta(t-\tau)-\theta(t))}\right)\hat{a}^\dagger(t)_{X1}\hat{b}^\dagger(t)_{X'0}\end{bmatrix}|00\rangle_s|00\rangle_i, \qquad (2)$$

where the global phase factor is omitted for simplicity. $|00\rangle_s$ and $|00\rangle_i$ denote the vacuum states at the two output ports (X0 and X1) and (X'0 and X'1). $\theta(t)$ denotes the relative phase of pump laser. $\theta_1$ and $\theta_2$ are the relative phase shifts between the long and short arms of the AMZIs at the signal and idler channels, respectively. The two-photon interference can be measured by changing the relative phase shifts between the long and short arms for the PLC-AMZI by controlling the operation temperature. The visibility of this interference is an essential value to estimate the time-bin entanglement. It exceeds the classical limit 70.1% only if the photons are entangled. As shown in our previous work [18], the visibility of the final state at Z-basis can be guessed precisely by using results shown Fig. 3. The final state at Z-basis is same result shown in Fig. 3 with the additional transmission loss is estimated as 8dB including intrinsic loss of 6dB. We obtain the visibilities of 98% at Z-basis. The following discussion will focus on X-basis outputs.

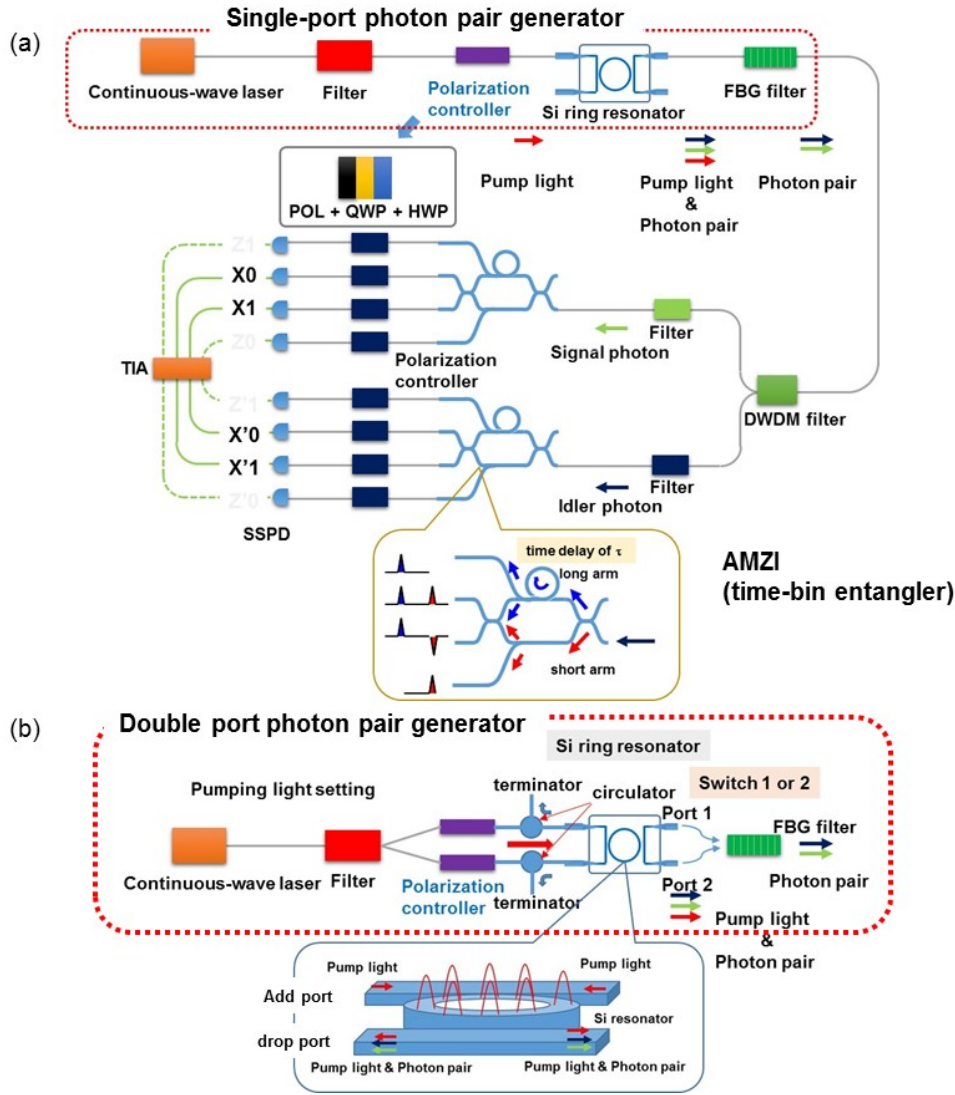

Fig. 4. (a) Conceptual view of experimental setup for time-bin entanglement, (b) spatial multiplexing of time-bin entanglement experiment with double coherent pump light input. Hatched area in (a) is replaced by (b).

*3.1 Entangled photon pair generation of the second pair*

In our previous work [18], time-bin entangled pair generation was confirmed using the first neighboring pair of the multiple resonant peaks in the wavelength domain. To substantiate wavelength division multiplexed entanglement, we have to measure visibilities for entangled photons in the second resonance peak pair, using the AMZIs. Figure 5(a) and 5(b) show the coincidence counts of X0(signal)-X'0(idler) (blue) and X0(signal)-X'1(idler) (red) for the second and first pairs in Fig. 2, respectively. The pump power is set to be 0.5 mW. We measure them for 60 or 300 seconds and three times for each plot, and the error bar is given by the standard deviation. Note that the size of the error bar in each point is smaller than the symbol size. The visibilities estimated by sinusoidal curve fitting are 82.22±2.22%/95.03±3.38% (X0-X'0/X0-X'1), and 81.96±3.16%/90.18±4.80% (X0-X'0/X0-X'1) respectively. The errors are due to the fitting error. These visibilities clearly exceed the value of the classical limit~71%

(1/√2) which concludes that our wavelength division multiplexed time-bin entangled photon pairs are entangled.

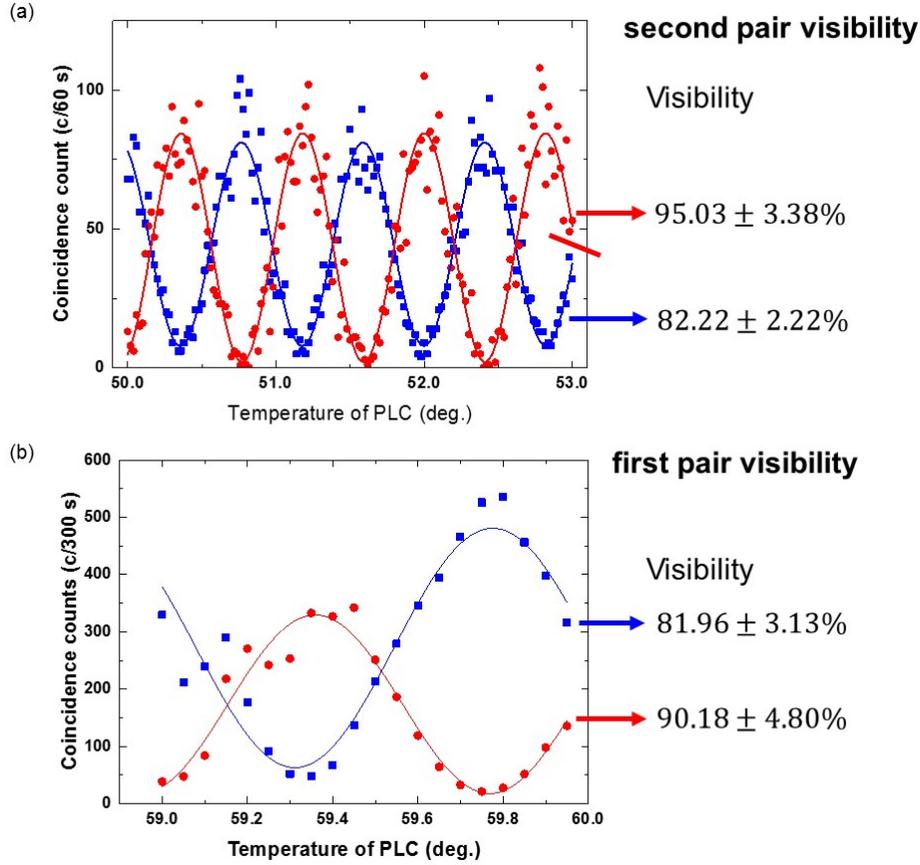

Fig. 5. X-basis coincidence counts of the second pair as a function of the operation temperature of the PLC (signal photon side). (a) Visibilities of the second pair, and (b) visibilities of the first pair in Fig. 2. Red points are coincidence counts of X0-X'0. Blue points are of X0-X'1. Fitted curves are obtained using sinusoidal function. The pump light power is set 0.5mW. The time window for one data point is set 64 ps.

*3.2 Demonstration of photon pair generation at two drop ports*

Our Si ring resonator has two input ports and two drop ports (see [Fig. 1]). To save the device resources, it might be useful if one can use these two drop ports simultaneously as separate entangled photon pair sources. Concerns of the double-port pumping in quantum application, however, are back scattering of spontaneous emission and some unwanted nonlinear phenomena [24] that could degrade the entanglement. The feasibility of the double-port pumping is tested by the setup described in Fig. 4(b). One pump light is divided into two, and these two pump light are input to the Si ring resonator in counter-propagating manner through polarization controllers and circulars. Each pump light enters the waveguide from both sides and is terminated at the third port of the circular. The visibilities are measured using the first pair. Figures 6(a) and 6(b) show the coincidence counts and visibilities of the first pair; 6(a) with the pump light from both side, and 6(b) at another drop port (port 2 in [Fig. 4(b)]) with the double-port pumping. All visibilities exceed the value of the classical limit. Comparing Fig. 6(a) with Fig. 5(b), the visibility with the double pumping is slightly better than that of the single pump light. Also, surprisingly, entangled photon pair generation rate at each drop port increases almost twice as that of the single-port pumping. The average photon pair generation rate by the single-port pumping (average of the two peak count rates in [Fig. 5(b)]) is 432 per 300 sec while the pair generation rate for the double-port pumping is 975 per 300 sec. The visibility and the pair generation rate at the other drop port (port 2) are plotted in Fig. 6(b) which also shows the rate increase as twice. Thus we conclude that our double-port pumping technique is useful not only to simply double the source by one device, but also to enhance each pair generation rate by some collaborative effect between two pump light.

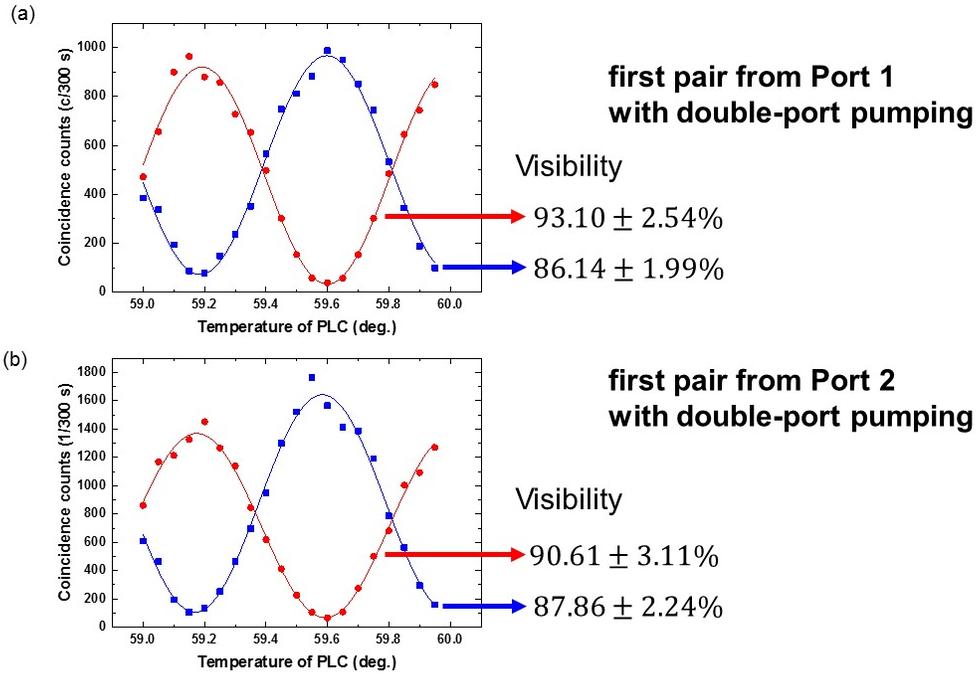

Fig. 6. X-basis coincidence counts of the first pair as a function of the operation temperature of the PLC (signal photon side) by the double-port pumping at (a) Port 1 and (b) Port 2. Red points are coincidence counts of X0-X'0. Blue points are of X0-X'1. The temperature of Si ring is 32.207°C. The pump light power from each input port is set 0.5 mW. The time window for one data point is set 64 ps.

*3.3 Discussion*

It is not clear the reason why the photon pair generation rate at each drop port is increased by the double-port pumping. To clarify the mechanism, we compare the transmission rate of the pump light observed at one drop port in Figs. 7 (a) with the single-port pumping, and (b) with the double-port pumping. While average transmission rates with 0.1 nm spectrum band with in both figures are almost the same, the output for the double-port pumping [Fig. 7(b)] shows a fringe-like structure and large fluctuations at the resonant wavelength. The error bars are defined as standard deviations of ten times measurements. Figure 7(b) strongly suggests that the second (counter propagating) pump light is partially reflected in the ring and the reflected light is interfered with the original pump light (note that the two pump light has the phase coherence). These interference fringe peaks of the pump light could contribute to increasing the average pumping power near the center of the resonance wavelength and raising the pair generation rate, since the SFWM is proportional to the square of the pumping power.

The question remained is where and why the pump reflection occurs. Optical impedance mismatch would take place at nearest points between Si waveguides and the ring resonator, and this impedance mismatch may cause reflection [25,26]. The point is why the reflection has been enhanced in double-port pumping. Possible cause is the optical effect that changes the state of the Si ring resonator. Thermo-optic effect [27] and carrier density modulation [28] could change the real part of the refractive index and may induce the reflection of the pump light in the resonator. In our experiment, the temperature of the Si ring resonator is controlled by using the Peltier cooler, but we do not use electrodes on the Si resonator. Thus it may be the case that the thermo-optic effect occurs at the local region in the Si ring resonator. Further investigation would be necessary to have a clear conclusion.

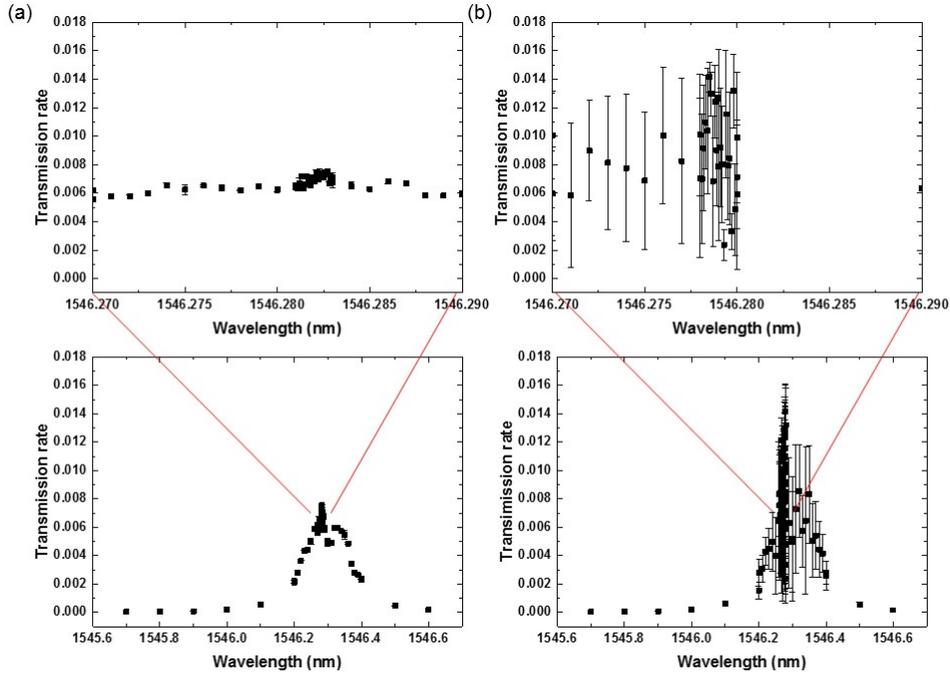

Fig. 7. Transmission rates from one add port to a drop port at 32.207 °C. (a) The pump laser is input from one add port. (b) The pump laser is input from two add ports. Error bars correspond to standard deviations of ten time measurements.

## 4. Summary

We demonstrate the wavelength multiplexed and double-port pumped generation of time-bin entangled photon pairs by using the Si ring resonator with 10 μm radius in the telecom C-band. The Si photonics integration and the wavelength multiplexing are important directions toward a compact and practical entangled photon source for future QKD and related quantum information processing applications. We also observe the SFWM enhancement by the double-port pumping technique. The double-port pumping technique can allow one to have two entanglement sources in one Si ring device and moreover enhance each pair generation rate by the collaborative effect between two pump light. To understand the enhancement mechanism better, we measure the transmission rate of the pump light and observe the fringe-like fluctuation. It strongly suggests that the pump light would be partially reflected by the local change of the refractive index of the resonator and the interference occurs between the main pump light and the reflected one. This interference causes large peaks of the pump electric field which contributes to enhance the photon pair generation rate. We have discussed the several possibilities of the origin of the reflection. It is an important future direction to have full understanding of the mechanism and use it to widen the functionality of Si ring resonators as quantum optical devices.